# Thermal Conductivity Mapping of Oxidized SiC/SiC Composites by Time-Domain Thermoreflectance with Heterodyne Detection

Xiaoyang Ji, Zhe Cheng, Ella Kartika Pek, David G. Cahill

Department of Materials Science and Engineering and Materials Research Laboratory,

University of Illinois at Urbana-Champaign, Urbana, IL 61801, USA

**Abstract:**

Silicon carbide/silicon carbide (SiC/SiC) composites are often used in oxidizing environments at high temperatures. Measurements of the thermal conductance of the oxide layer provide a way to better understand the oxidation process with high spatial resolution. We use time-domain thermoreflectance (TDTR) to map the thermal conductance of the oxide layer and the thermal conductivity of the SiC/SiC composite with a spatial resolution of 3 μm. Heterodyne detection using a 50-kHz-modulated probe beam and a 10-MHz-modulated pump suppresses the coherent pick-up and enables faster data acquisition than what has previously been possible using sequential demodulation. By analyzing the noise of the measured signals, we find that in the limit of small integration time constants or low laser powers, the dominant source of noise is the input noise of the preamplifier. The thermal conductance of the oxide that forms on the fiber region is lower than the oxide on the matrix due to small differences in thickness and thermal conductivity.

## 1. Introduction

Silicon carbide (SiC)-based composites are widely used in advanced engines, heat exchangers, gas turbines, and equipment for materials processing at high temperatures.[1] SiC-fiber-reinforced SiC matrix composites (SiC/SiC composites) are also being investigated for potential applications in nuclear fusion reactors due to their low activation, radiation resistance, and relatively low neutron absorption.[2]

Many of the current and potential applications of SiC/SiC composites involve oxidizing atmospheres.[3,4] The presence of water vapor accelerates oxidation.[5–8] In $H_2O/O_2$ gas mixtures at 1200 °C to 1400 °C, SiC reacts with water vapor and forms volatile $Si(OH)_4$, leading to material degradation.[4,9] Moreover, the SiC matrix and SiC fibers show slightly different oxidation behaviors. Cristobalite can be observed along with amorphous $SiO_2$ during the oxidation, but the temperature of cristobalite formation for the fibers is lower than that for the matrix.[10] For example, the cristobalite is detected at 1100 °C for Si-C-O Nicalon fibers in dry oxygen and 1400 °C for Hi-Nicalon fibers in dry oxygen/nitrogen atmosphere, while the temperature at which the cristobalite is found for SiC-matrix is 1500 °C in dry oxygen.[10] The oxidation of fibers usually has a higher rate than bulk SiC or single crystal SiC materials due to the presence of impurities (Cl, S, Ca, etc.) and surface curvature effects.[11]

Time-domain thermoreflectance (TDTR) can map spatial variations in the thermal conductivity of materials with micron-scale spatial resolution. Huxtable *et al.* reported thermal conductivity mapping over large areas by collecting the TDTR ratio signal at a fixed delay time.[12] The micrometer-scale resolution of TDTR mapping enabled the detection of features such as grain boundaries,[13,14] composition distribution,[15–18] the quality of mechanical joints,[19] and interface conditions.[20,21]

In this work, we improve the speed of TDTR mapping by replacing the conventional sequential demodulation of the TDTR signal with heterodyne detection. In the conventional approach, the intensity of the probe beam is modulated at 200 Hz by a mechanical chopper. The outputs of a radio-frequency (RF) lock-in that is synchronized to the 10 MHz modulation frequency of the pump is measured by two audio-frequency (AF) lock-ins synchronized to the 200 Hz modulation of the probe. In the heterodyne approach, we use a photoelastic modulator (PEM) operating at 50 kHz to modulate the probe intensity. The signal is detected with a rf lock-in that referenced to the frequency of the sum of the modulation frequencies of the pump and the probe beam.

We apply the heterodyne TDTR mapping technique to map spatial variations in two unknown thermophysical properties of an oxidized SiC/SiC composite: the thermal conductivity of the composite and the thermal conductance of the thin oxide layer that forms on the surface of the composite by oxidation in dry air at 1000°C. The determination of the two unknown parameters requires two independent measurements, e.g., the TDTR signals collected at two delay times. The two values of the delay times are chosen to optimize the difference in sensitivities. We use a lookup table of calculated values of the TDTR ratio signals to efficiently determine the thermal conductance and the thermal conductivity at each location in a map of the thermal properties.

**2. Experimental Methods**

*1. Sample fabrication*

The SiC/SiC composite we studied was fabricated by HyperTherm, Inc., and is comprised of Hi-Nicalon Type S fibers (~10 μm diameter), a chemically vapor-infiltrated (CVI) matrix, and interphase materials coated on the fibers.[22] The fibers are aligned in 2-D cross bundles.[22] The interphase material between the fibers and the matrix is comprised of multilayers of pyrolytic carbon (PyC) and SiC (one 40-nm PyC layer and four periods of 50 nm SiC/10 nm PyC).[23] The SiC/SiC composite was mechanically polished using Allied Multiprep with increasingly fine grades of diamond polishing pads (3 μm, 1 μm, 0.5 μm, and 0.1 μm) for ~1 h for each diamond polishing pad. The sample was then cleaned by acetone and ion-beam polished using a Gatan PECS-II for 1 h. The ion beam polishing was done with a gun angle of 6°, an ion beam energy of 6 keV, and an ion beam current of ~20 mA.

*2. Oxidation of SiC wafers and SiC/SiC composite in dry air atmosphere*

We first oxidized single crystal SiC wafers (semi-insulating 4H-SiC, orientation <0001>) with different heating times and temperatures to determine the optimal conditions for oxidation of the SiC/SiC composite for the TDTR mapping measurements. SiC wafers were purchased from TankeBlue Semiconductor Co., Ltd. The SiC wafers were oxidized in dry air (7 ppm $H_2O$) in a tube furnace with a gas flow rate of 200 cc/min. The heating rate was 10 °C/min. The samples were transferred into the hot zone of the furnace when the aimed temperature (1000 °C) was reached. After heating, the samples were cooled naturally to room temperature by removing the sample from the hot zone.

We measured the thicknesses ($d$) of the oxide layers by X-ray reflectivity (XRR) using a Bruker D8 Advance diffractometer with Cu $K\alpha_1$ radiation; we fit the data to a $SiO_2$/SiC model using

LEPTOS software. Picosecond acoustics provides spatially resolved measurements of transit times of acoustic waves that reflect from the oxide/SiC interface. To perform picosecond acoustic measurements, the oxidized sample was coated with an Al film. A sub-picosecond laser pulse heats the metal film and launches a longitudinal acoustic wave.[24] The transit time $\Delta t$ between the acoustic echoes corresponding to the Al/oxide interface and the oxide/SiC interface was recorded. The relation of thickness $d$ and transit time $\Delta t$ was fitted linearly.

Given the results of SiC wafers, we heated the SiC/SiC composite at 1000 °C for 2 h in dry air in the tube furnace. The thickness of the oxide formed in 2 h allows us to determine the thermal properties of both the oxide and the SiC/SiC composite with good sensitivities. The oxidized SiC/SiC composite was coated with a layer of Al (~ 80 nm) by dc magnetron sputtering. The Al film serves as a transducer for picosecond acoustics and TDTR measurements. Figure 1(a) shows a dark-field optical image of the surface of the oxidized SiC/SiC composite coated with Al, captured by a charge-coupled device (CCD) camera in the TDTR apparatus, using a 20× objective lens with numerical aperture of 0.42. On the left side of the area shown in Figure 1(a), the fibers are perpendicular to the sample surface; on the right side of the area shown in Figure 1(a), the fibers are parallel to the sample surface. In this dark field optical microscopy image, the matrix is darker than the fibers. Figure 1(b) shows the schematic diagram of the structure of the SiC/SiC composite.

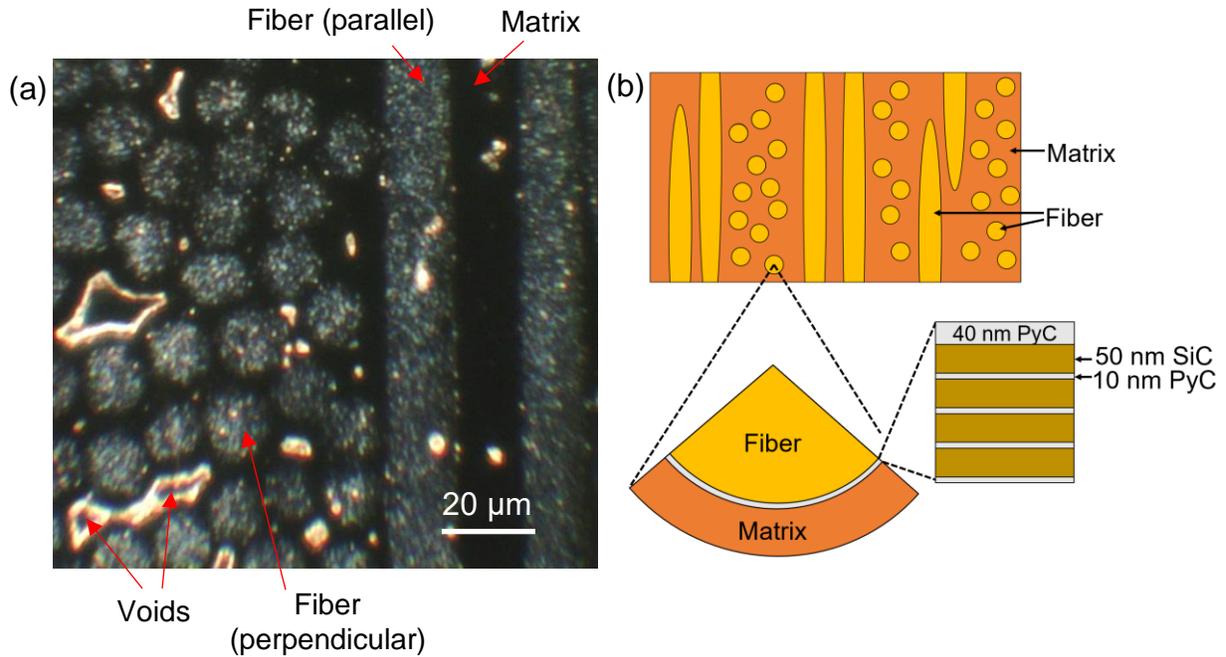

Figure 1. (a) Dark-field optical image of the surface of the oxidized SiC/SiC composite coated with Al. (b) Schematic diagram of the structure of the SiC/SiC composite. The SiC/SiC composite we studied is composed of ~10 μm diameter Hi-Nicalon Type S fibers, chemically vapor-infiltrated (CVI) matrix, and interphase materials. The interphase materials between the fibers and the matrix are pyrolytic carbon (PyC) SiC multilayers. The schematic diagram is not scaled proportionally.

*3. Thermal conductivity measurements*

*3.1 Conventional and heterodyne time-domain thermoreflectance (TDTR)*

The TDTR apparatus has been described previously.[25] A mode-locked Ti:sapphire laser (~785 nm) is split into a pump beam and a probe beam using a "two-tint" approach with sharp-edged optical filters.[26] The pump beam produces a temperature jump on the sample surface; the probe beam detects changes in the temperature of the sample surface through changes in optical reflectivity, i.e., thermoreflectance.[27] In our conventional TDTR approach, using sequential demodulation of the thermoreflectance signal, the intensity of the pump beam is modulated by an electro-optic modulator (EOM) at a frequency near 10 MHz and the intensity of the probe beam is modulated by a mechanical chopper at a frequency near 200 Hz. The reflected probe beam is picked up by a photodiode and a rf lock-in amplifier synchronized to the modulation frequency of the pump. The in-phase and quadrature outputs of the rf lock-in are then measured by two computer-based lock-ins that are synchronized to the modulation frequency of the probe.[28] This double-modulation scheme is important to remove coherent pickup.[28] We typically choose a time constant for the audio-frequency lock-ins of 700 ms in the conventional TDTR measurements. As we show below, in the conventional approach using a 200-Hz modulation frequency of the probe, the minimum usable integration time constant for the two audio-frequency lockins is on the order of 30 ms.

In our heterodyne TDTR approach, described for the first time here, the intensity of the probe beam is modulated by a photoelastic modulator (HINDS Instruments, PEM 100) at a frequency of 50 kHz. We operate the PEM so that the PEM is equivalent to a quarter-wave plate at the maximum and minimum of the phase modulation.[29] The PEM is placed between a half-wave plate that controls the incident angle of polarization and a quarter-wave plate that converts the oscillating elliptical polarization to an oscillating linear polarization. After the quarter-wave plate, the polarization of the probe beam oscillates between horizontal and vertical. Since the probe passes through a polarizing beam before reaching the sample, the rotating polarization of the probe is converted to an intensity modulation with fixed horizontal polarization.

The signals at the sum frequency (10.7 MHz) of the modulation frequency of the pump (10.65 MHz) and the modulation frequency of the probe (50 kHz) are measured by a rf lock-in amplifier.

We use a mixer (Mini-Circuits, ZAD-6+) to generate a sum and difference frequency signal. A quartz crystal filter (DP-iot crystal filter AM filter 10.7 MHz±7 kHz) passes only the sum frequency (10.7 MHz) to the reference channel of the rf lock-in. Figure 2 summarizes the electrical connections. In addition to the thermoreflectance signals measured by the rf lock-in amplifier, the reflected probe beam is also measured by an audio-frequency lock-in amplifier synchronized to the 50 kHz modulation frequency of the PEM to measure the reflected intensity of the probe beam.

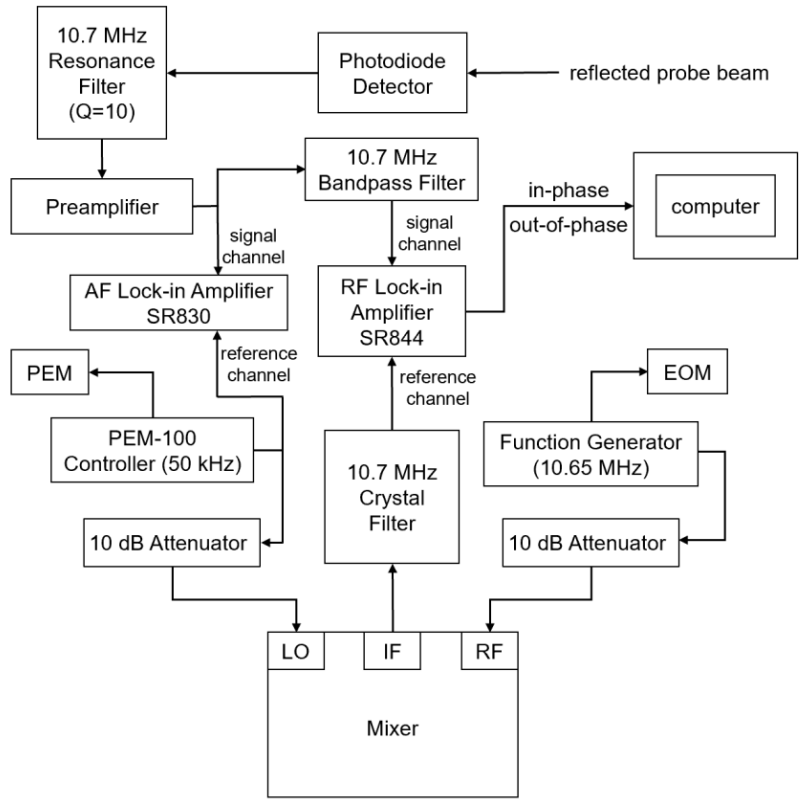

Figure 2. Schematic diagram of the heterodyne configuration of the TDTR measurement. The 10.7 MHz Q=10 resonance filter is an inductor in series between the photodiode detector and the 50 ohm input impedance of the rf preamplifier. AF: audio frequency; RF: radio frequency; PEM: photoelastic modulator; EOM: electro-optic modulator; LO: local oscillator; IF: intermediate frequency.

*3.2 Noise measurements*

We evaluated noise in the conventional and heterodyne TDTR measurements as a function of the integration time constants $\tau$ over the range 100 µs$\leq \tau \leq$1 s using a SiO$_2$(500 nm)/Si reference sample coated with a 70-nm Al layer. A 20× (1/e$^2$ radius $w_0$=2.7 µm) objective lens was used in the noise measurements.

To determine the noise, the TDTR signals were measured 400 times at a single spot on the reference sample. The noise is calculated as a root-mean-square deviation of the signals. The time between the measurements, i.e., the dwell time $t$, was set to be the integration time constant except when the time constant ($\tau$) is shorter than 1 ms. When the time constant is shorter than 1 ms, the dwell time was fixed at 1 ms. The noise of the conventional TDTR was also analyzed for comparison, where the dwell time was the same as the time constant over the range 1 ms$\leq \tau \leq$1 s (except when $\tau$=700 ms, we chose $t$=500 ms).

*3.3 Thermal conductivity mapping of the SiC/SiC composite*

We mapped the oxidized SiC/SiC composite by both the conventional TDTR and the heterodyne TDTR approaches with the 20× objective lens and a piezoelectric two-dimensional positioning stage (Physik Instrumente, Q-545.240). We scanned a 800 µm$^2$ (40 µm×20 µm) area of the composite sample with a step size of 1 µm at a pump-probe delay time of 200 ps and then repeated the mapping measurement with a pump-probe delay time of 3.6 ns.

In the measurement process, the stage moves to a new measurement position and waits for a dwell time ($t$) before the signals are recorded ($V_\text{in}$: in-phase voltage, $V_\text{out}$: out-of-phase voltage, ratio signal: $-V_{in}/V_{out}$). The stage has a closed loop control system. At the end of a row, the stage moves to the end of the next row. The stage then moves to the beginning of that row, and collects the data for that row. The waiting time at the beginning of the new row is twice the dwell time of each mapping step. In the conventional TDTR mapping approach, the time constant was 700 ms and the dwell time was 500 ms, corresponding to a scanning speed of 2 μm/s. In the heterodyne TDTR mapping, we did two sets of measurements using time constants of 100 ms and 30 ms. These time contants were chosen as the optimal trade-off between signal-to-noise and the time required to acquire the map of thermal properties. The dwell times were set to be the same as the time constants, corresponding to scanning speeds of 10 μm/s and 33 μm/s.

The unknown thermal parameters are obtained by fitting the measured TDTR ratio to an analytical solution of the heat transfer equation for a model of the sample structure.[30] We used a three-layer model (Al/oxide/composite). Literature values were used for the heat capacities of Al (2.43 J/cm$^3$-K), the oxide (1.62 J/cm$^3$-K), and the SiC/SiC composite (2.17 J/cm$^3$-K).[17,31,32] The unknown parameters were the effective thermal conductance of the oxide layer and the thermal conductivity of the SiC/SiC composite.

Since the oxide layer was thin, the measurement cannot distinguish well between the thermal resistances of the interfaces and the thermal resistance of the oxide layer; i.e., the measurement is mostly sensitive to the series thermal conductance ($G_{oxide}$) of the oxide and the two adjacent interfaces:

$$\frac{1}{G_{oxide}} = \frac{1}{G_{Al\text{-}oxide}} + \frac{d}{\Lambda_{oxide}} + \frac{1}{G_{oxide\text{-}SiC}}, \tag{1}$$

where $G$ is the thermal interface conductance of the Al/oxide or oxide/composite interfaces, $d$ is the thickness of the oxide layer, and $\Lambda_{oxide}$ is the thermal conductivity of the oxide.[33,34] In the analysis of the data presented below, we fixed the thickness of the oxide layer as $d=30$ nm, the approximate average thickness of the oxide layer on the composite sample as discussed in the following part.

To determine the two unknown thermal properties, we mapped the TDTR signals with two delay times. The two delay times were chosen based on the TDTR sensitivities.[35] The TDTR sensitivities are defined by

$$S_\alpha = \frac{d\ln(-V_{in}/V_{out})}{d\ln\alpha}, \qquad (2)$$

where $\alpha$ is any of the parameters in the sample structure.[36] $G_{oxide}$ and $\Lambda_{SiC}$ were first measured using the heterodyne TDTR data acquired over a full scan of the pump-probe delay time. Then the sensitivities to $G_{oxide}$ and $\Lambda_{SiC}$ were calculated for the typical properties of the fiber and matrix. $\Lambda_{SiC} = 17$ W/m-K for the fiber and $\Lambda_{SiC} = 57$ W/m-K for the matrix. Typical values of the thermal conductance of the oxide were $G_{oxide} = 33$ MW/m²-K for the fiber and 41 MW/m²-K for the matrix. The sensitivities are summarized in Figure 3. The sensitivities to $G_{oxide}$ are high at short delay times and low at long delay times while the sensitivities to the thermal conductivities of the fiber and matrix are nearly independent of delay time.

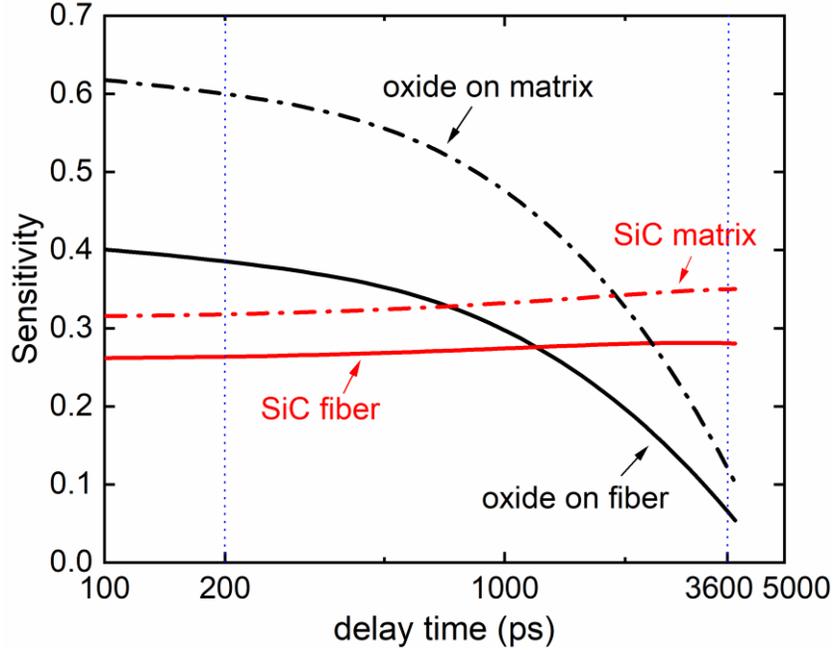

Figure 3. The sensitivities of $G_{oxide}$ and $\Lambda_{SiC}$ in heterodyne TDTR for typical regions of the SiC fiber (solid lines) and the SiC matrix (dash dot lines). The three-layer structure (Al/oxide/composite, Al: 86 nm; oxide: 30 nm) was used for calculation. The $1/e^2$ radius of the laser was 2.7 μm.

To analyze the TDTR ratios at the two delay times, we created a two-dimensional table of calculated ratio values for a range of $G_{oxide}$ and $\Lambda_{SiC}$ values:[30] 20< $G_{oxide}$ <60 MW/m$^2$-K and 10< $\Lambda_{SiC}$ <100 W/m-K. We picked 200 linearly distributed values within both the two ranges; i.e., we calculated 40,000 entries in the table. At each pixel in the mapping data, we compared the experimental ratio signals acquired at delay times of 200 ps and 3600 ps to the table of theoretical ratios to determine $G_{oxide}$ and $\Lambda_{SiC}$. The calculation was completed in ~30 minutes using an uncompiled Matlab script.

**3. Results and discussion**

*1. Oxidation of the SiC/SiC composite*

The oxide thicknesses *d* and the corresponding transit times Δ*t* are plotted in Figure 4 of SiC wafers oxidized at 1000 °C for 2, 4, and 8 h. The dashed line is the linear fitting of data points. The thicknesses of the oxide layer formed on the SiC/SiC composite at randomly selected locations were determined based on this linear relationship between the positions of the acoustic echoes and thickness. The average thicknesses of the oxide layer on the fibers perpendicular and parallel to the surface are similar, 33 nm and 35 nm, respectively. The matrix has an average thickness of 25 nm, significantly thinner than the oxide layer on the fibers. Moreover, the thickness of the oxide layer on fibers also has a larger standard deviation than the matrix, 7 nm and 1.3 nm, respectively. The oxide formed on the SiC/SiC composite is thicker than the oxide formed on the SiC wafer for this oxidation condition of 1000 °C for 2 h in dry air.

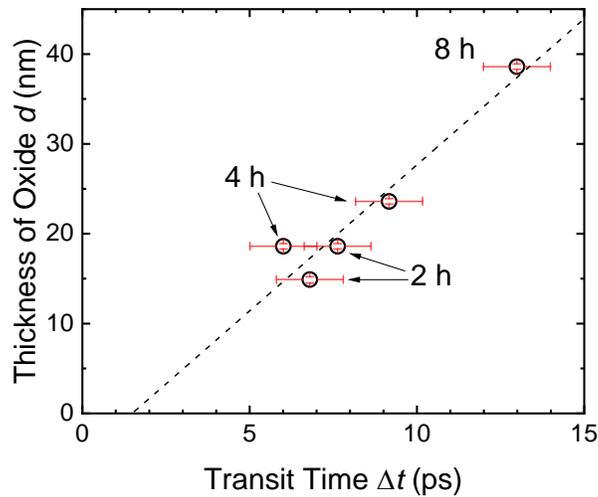

Figure 4. Thickness *d* determined by x-ray reflectivity plotted as a function of the acoustic round-trip transit time Δ*t* of the oxide on SiC wafers which were oxidized at 1000 °C for 2, 4, and 8 h in dry air. The error bars on the data points denote the uncertainty in the measurements of each sample.

*2. Noise analysis*

The noise in the TDTR signals derives from the noise of the electronics and intensity noise of the laser. A key conclusion of our analysis summarized below is that for short time constants and low laser powers, the noise of the TDTR signals is dominated by the input noise of the SR445A preamplifier. The input noise specified by the manufacturer is 6.4 nV/√Hz. The gain of the preamp is 5. Therefore, in the limit of short time constants and low laser powers, the noise in the TDTR signals, $\Delta V_{rms}$, is

$$\Delta V_{rms} = \frac{32 \text{ nV}}{\sqrt{\text{Hz}}} \sqrt{\frac{1}{4\tau}} = \frac{16 \text{ nV}}{\sqrt{\text{Hz}}} \frac{1}{\sqrt{\tau}},$$

where $\tau$ is the time constant of the rf lock-in amplifier, and $1/(4\tau)$ is the equivalent noise bandwidth.[37]

In addition to the noise from the preamplifier, intensity fluctuations of the laser contribute to the noise. The laser intensity noise can be divided into two regimes. The laser intensity noise near the modulation frequency makes a contribution to the noise of the TDTR signals that scales linearly with probe power. The laser intensity noise at low frequencies, i.e., frequencies on the order of the data acquisition rate, makes a contribution to the noise of the TDTR signals that scales with the product of the pump and probe power.

The noise data are summarized in Figure 5. When the laser is off or only the probe beam arrives at the sample surface, the noise scales inversely with the square root of the integration time constant, $1/\sqrt{\tau}$, and is comparable to but slightly lower than the input noise of the preamplifier. With both the pump and probe beams incident on the sample, the noise is essentially unchanged at small time constants, but the noise deviates from the $1/\sqrt{\tau}$ trend for long time constants. In this regime, the noise has a strong dependence on laser power, with a scaling that approximately follows the product of the pump and probe powers. These results suggest that the dominant noise at low laser powers and short time constants is the input noise of the rf preamplifier and that the dominant noise at high laser powers and long time constants is the low frequency intensity noise of the laser.

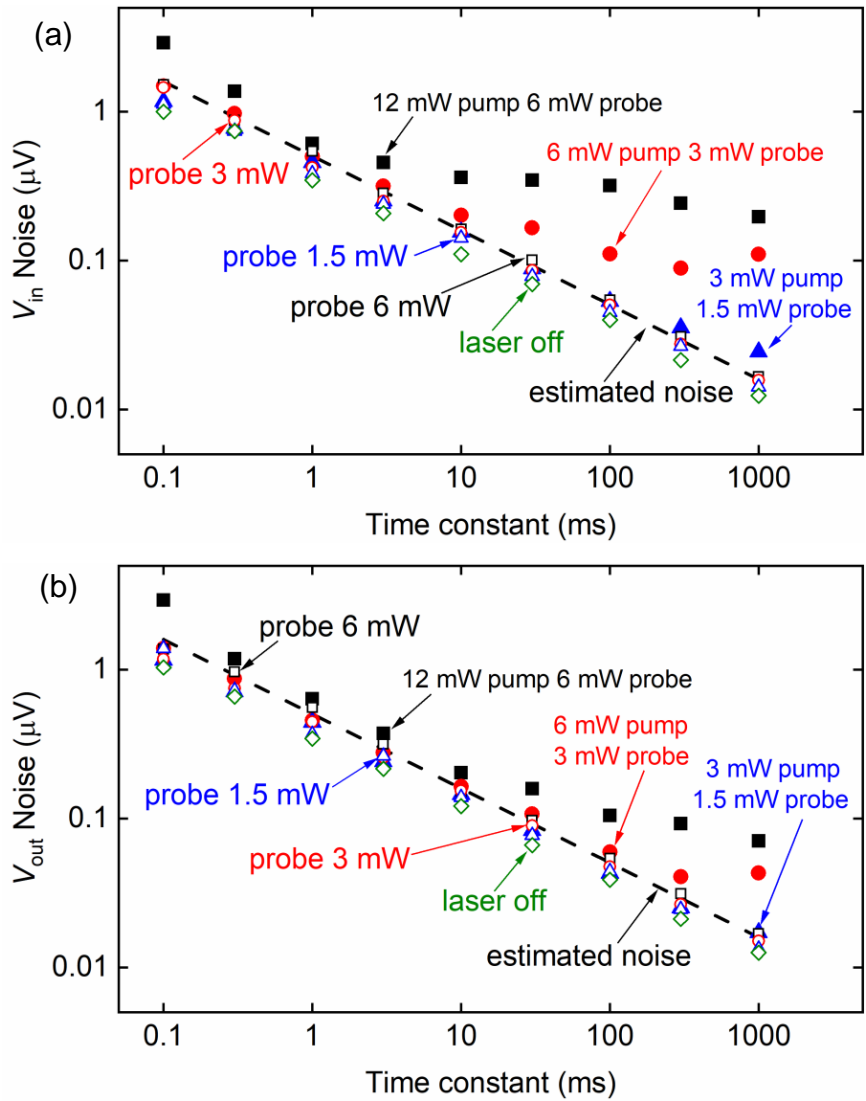

Figure 5. (a) The noise of in-phase voltage signals in the heterodyne setup with different laser beams; (b) The noise of out-of-phase signals in the heterodyne setup with different laser beams: 12 mW pump and 6 mW probe beams (filled square), 6 mW pump and 3 mW probe beams (filled circle), 3 mW pump and 1.5 mW probe beams (filled triangle), only 6 mW probe beam (open square), 3 mW probe beam (open circle), 1.5 mW probe beam (open triangle), and laser off (open diamond). The dash lines are the estimated noise calculated from the noise of the preamplifier. The 20× objective lens was used in these measurements.

TDTR measurements of thermal transport properties, however, depend on the ratio signal ($-V_{in}/V_{out}$) and not the separate in-phase and out-of-phase signals. The noise of the ratio signal are summarized in Figure 6. The ratio signal is significantly less effected by low frequency intensity noise in the laser because both the in-phase and out-of-phase signals have the same dependence on laser power.

The data plotted in Figure 6 also compare the noise of the heterodyne TDTR approach to the noise of the conventional TDTR approach. The noise of the ratio signals are similar for the two approaches at long time constants. However, the noise of the conventional TDTR signals increases dramatically when the time constant is less than 30 ms due to inadequate averaging over the 200 Hz modulation frequency of the probe.

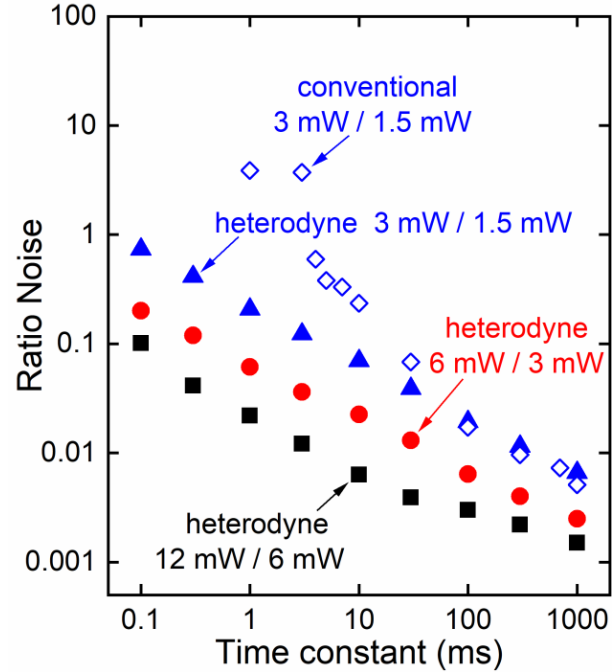

Figure 6. Comparison of the noise of ratio (-$V_{in}$/$V_{out}$) for the conventional TDTR (open diamond) and the heterodyne TDTR setup with 12 mW pump and 6 mW probe (filled square, 12 mW/6 mW), 6 mW pump and 3 mW probe (filled circle, 6 mW/3 mW), and 3 mW pump and 1.5 mW probe power (filled triangle, 3 mW/1.5 mW). The 20× objective lens was used in these measurements.

*3. Thermal conductivity mapping of the oxidized SiC/SiC composite*

The mapping area on the oxidized SiC/SiC composite was chosen to include the fiber, the matrix, and the interphase, as shown in Figure 7(a). Figure 7(b) shows the mapped thermal conductance of the oxide layer by the conventional TDTR setup with scanning speed of 2 μm/s. Figure 7(c) and 7(d) show the thermal conductance of the oxide layer mapped by the heterodyne TDTR setup with scanning speeds of 10 μm/s and 33 μm/s, respectively. The profiles of the fibers (both perpendicular and parallel to the surface) and the matrix are clear, consistent with the CCD camera image. Thus, the mapping techniques with high spatial resolution can capture the structural features (the fiber and the matrix). When the scanning speed (10 μm/s) is 5 times higher than the conventional TDTR mapping, the images are clear and the boundaries between the fibers and the matrix are easy to discern. When the scanning speed (33 μm/s) is 17 times faster than the conventional TDTR setup, the mapped image loses details and cannot capture the small structural features. The increased scanning speed leads to increased noise that degrades the image quality.

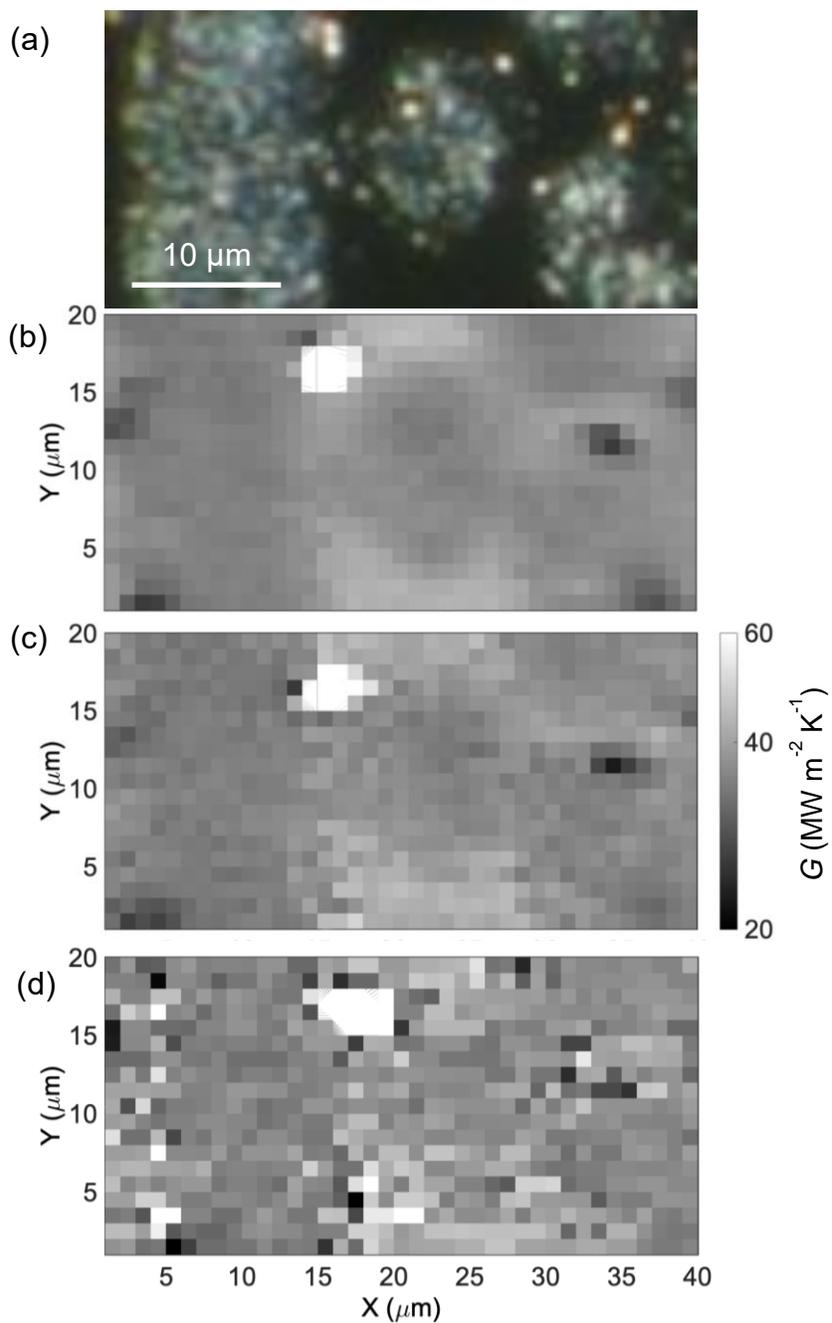

Figure 7. (a) The scanning area captured by the CCD camera. This image of the SiC/SiC is included as a reference for the maps of the oxide layer thermal conductance shown in panels (b)-(d). (b) scanning speed of 2 μm/s using the conventional TDTR setup; (c) scanning speed of 10 μm/s using the heterodyne TDTR setup; (d) scanning speed of 33 μm/s using the heterodyne TDTR setup. The powers of the pump and probe beam were 8 mW and 4 mW, respectively. The 20× (1/$e^2$ radius $w_0$=2.7 μm) objective lens was used in these measurements.

The effective thermal conductance of the oxide that forms on the SiC fibers are in the range 27~40 MW/m$^2$-K, approximately 20% lower than the oxide that forms on the matrix (40~47 MW/m$^2$-K). The Hi-Nicalon Type S fibers are composed of β-phase SiC grains with widely varying grain sizes (tens to hundreds of nanometers) and excess carbon in multi-grain junctions.[38] The SiOC glass pockets in SiC fibers, partially occupied with turbostratic graphite, may result in the graphite residue beneath the oxide/SiC interface.[11] Accordingly, the oxide layer on fibers has a lower thermal conductance because of more complicated and defective microstructures compared to the matrix. In addition, the average thickness of the oxide layer on the fibers are slightly larger than those on the matrix.

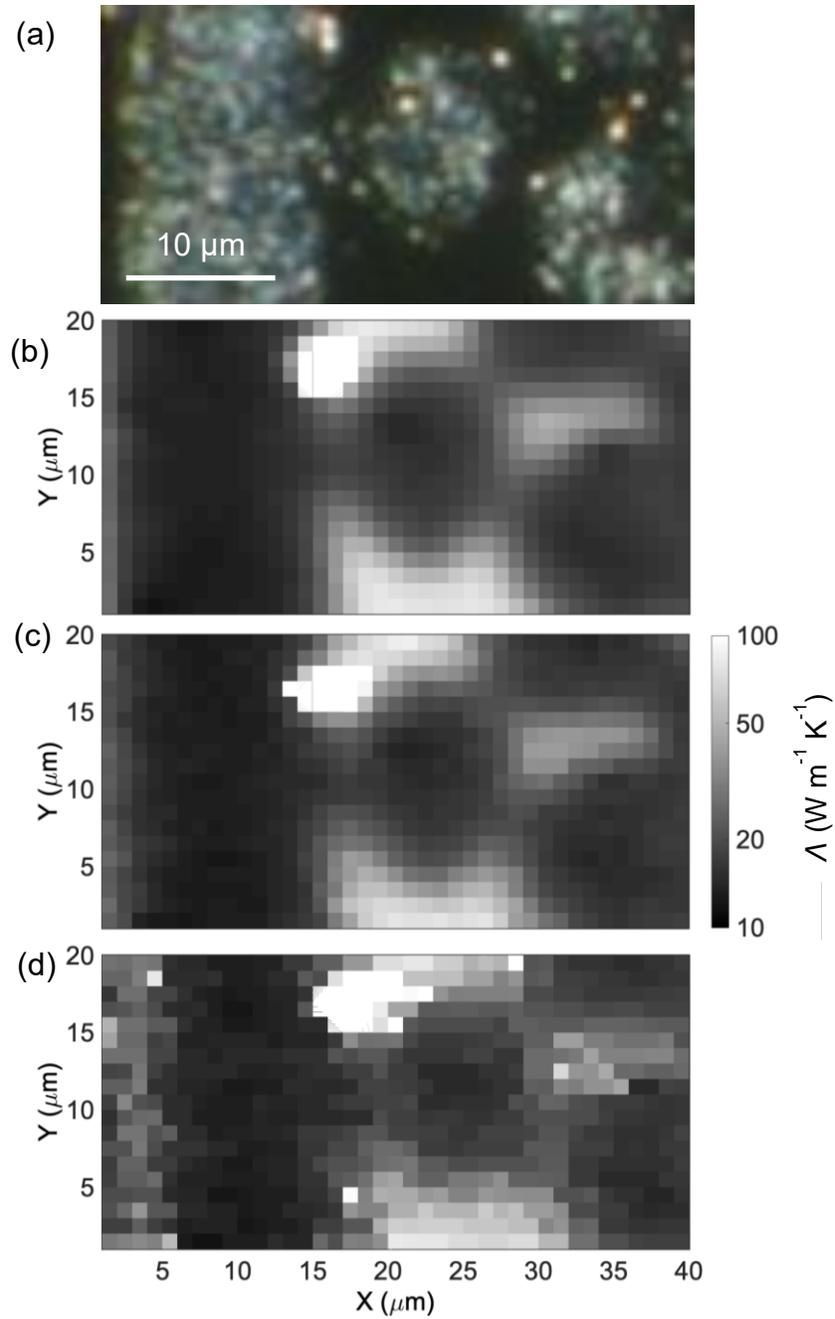

Figure 8. (a) The scanning area captured by the CCD camera. This image of the SiC/SiC composite is essentially the same as Figure 7(a) and is included to as a reference for the thermal conductivity maps in panels (b)-(d). (b) scanning speed of 2 μm/s using the conventional TDTR setup; (c) scanning speed of 10 μm/s using the heterodyne TDTR setup; (d) scanning speed of 33 μm/s using the heterodyne TDTR setup.

Figure 8(b) shows the thermal conductivity maps of the SiC/SiC composite by the conventional TDTR setup with a scanning speed of 2 μm/s while Figure 8(c-d) show those mapped by the heterodyne TDTR setup with scanning speeds of 10 μm/s and 33 μm/s. The thermal conductivities of the matrix are within a range of 50~90 W/m-K. This is consistent with Ella Pek's result, a range of 50~120 W/m-K.[22] The thermal conductivities within the fiber regions are relatively uniform with an average value of 15 W/m-K. The thermal conductivity of fibers parallel to the surface (14 W/m-K) is slightly lower than that of fibers perpendicular to the surface (16 W/m-K). Ella Pek's work also showed some lower thermal conductivity (20 W/m-K) of the fibers parallel to the surface than the fibers in the perpendicular direction (23 W/m-K), but mostly fibers are uniform regardless of the direction with an average thermal conductivity of 22 W/m-K. The thermal conductivity of fibers in this work is ~32% lower than the previous work, which is probably due to the irreversible degradation of fibers during the high-temperature oxidation. The area near the edge of fibers or between fibers and fibers show a variation of thermal conductivity from around 18 to 50 W/m-K. The thermal conductivity in such area increases as the distance away from fibers increases, which was also observed in the previous work.[22]

## 4. Conclusions

In this work, we developed a heterodyne TDTR apparatus to accelerate the data acquisition rate in mapping of thermal properties. In the limit of short integration time constants and low laser powers, the noise of the heterodyne TDTR signals is dominated by the input noise of the rf preamplifier, while the low frequency intensity noise of the laser dominates at long time constants and high laser powers. Use of the ratio ($-V_{in}/V_{out}$) signal suppresses the contribution from the low frequency intensity noise of the laser. The heterodyne TDTR has a smaller noise at time constants of $\tau<100$ ms, which allows for a faster data collection than the conventional TDTR.

The mapped thermal conductivities keep the accuracy until a scanning speed of 33 μm/s in the heterodyne TDTR, a 17 times higher speed than the conventional TDTR mapping. The mapped thermal conductivity of the fibers was 15 W/m-K while that of the matrix was 50~90 W/m-K. The area near fibers has a medium thermal conductivity of 18~50 W/m-K. The effective thermal conductance of the oxide layer on the fibers was 27~40 MW/m$^2$-K while that of the oxide layer on the matrix was 40~47 MW/m$^2$-K. The differences of the thermal conductance of oxide layers on fibers and matrix are attributed to the different structures of the oxide layers on the fibers and the matrix. The higher thermal conductance of the oxide layer on the matrix than the fibers possibly results from the less defective oxide composition and slightly smaller thickness than the fibers.

**Acknowledgements**

The authors acknowledge financial support from the US Army CERL W9132T-19-2-0008. TDTR data and XRR data were obtained in the Materials Research Laboratory (MRL) at the University of Illinois at Urbana-Champaign. The authors thank Juan Sebastian Lopez for the X-ray reflectivity measurements and Darshan Chalise for his assistance with the heterodyne TDTR apparatus.

# REFERENCES

1. Naslain, R. Design, preparation and properties of non-oxide CMCs for application in engines and nuclear reactors: An overview. *Composites Science and Technology* vol. 64 155–170 (2004).

2. Koyanagi, T. *et al.* Recent progress in the development of SiC composites for nuclear fusion applications. *Journal of Nuclear Materials* vol. 511 544–555 (2018).

3. Watanabe, H. & Hosoi, T. Fundamental Aspects of Silicon Carbide Oxidation. in *Physics and Technology of Silicon Carbide Devices* (InTech, 2012). doi:10.5772/51514.

4. Terrani, K. A. *et al.* Silicon carbide oxidation in steam up to 2 MPa. *J. Am. Ceram. Soc.* **97**, 2331–2352 (2014).

5. Roy, J., Chandra, S., Das, S. & Maitra, S. Oxidation behaviour of silicon carbide - A review. *Rev. Adv. Mater. Sci.* **38**, 29–39 (2014).

6. Opila, E. J. Variation of the oxidation rate of silicon carbide with water-vapor pressure. *J. Am. Ceram. Soc.* **82**, 625–636 (1999).

7. Mogilevsky, P., Boakye, E. E., Hay, R. S., Welter, J. & Kerans, R. J. Monazite Coatings on SiC Fibers II: Oxidation Protection. *J. Am. Ceram. Soc.* **89**, 3481–3490 (2006).

8. Hay, R. S. & Corns, R. Passive oxidation kinetics for glass and cristobalite formation on Hi-Nicalon$^{TM}$-S SiC fibers in steam. *J. Am. Ceram. Soc.* **101**, 5241–5256 (2018).

9. Opila, E. J. & Hann, R. E. Paralinear oxidation of CVD SiC in water vapor. *J. Am. Ceram. Soc.* **80**, 197–205 (1997).

10. Naslain, R. *et al.* Oxidation mechanisms and kinetics of SiC-matrix composites and their constituents. in *Journal of Materials Science* vol. 39 7303–7316 (Springer, 2004).

11. Hay, R. S. & Chater, R. J. Oxidation kinetics strength of Hi-Nicalon $^{TM}$ -S SiC fiber after

oxidation in dry and wet air. *J. Am. Ceram. Soc.* **100**, 4110–4130 (2017).

12. Huxtable, S., Cahill, D. G., Fauconnier, V., White, J. O. & Zhao, J. C. Thermal conductivity imaging at micrometre-scale resolution for combinatorial studies of materials. *Nat. Mater.* **3**, 298–301 (2004).

13. Sood, A. *et al.* Direct Visualization of Thermal Conductivity Suppression Due to Enhanced Phonon Scattering Near Individual Grain Boundaries. *Nano Lett.* **18**, 3466–3472 (2018).

14. Cheng, Z. *et al.* PROBING LOCAL THERMAL CONDUCTIVITY VARIATIONS IN CVD DIAMOND WITH LARGE GRAINS BY TIME-DOMAIN THERMOREFLECTANCE. in *International Heat Transfer Conference 16* 8694–8701 (Begellhouse, 2018). doi:10.1615/IHTC16.tpm.022782.

15. Zhao, J. C., Zheng, X. & Cahill, D. G. Thermal conductivity mapping of the Ni-Al system and the beta-NiAl phase in the Ni-Al-Cr system. *Scr. Mater.* **66**, 935–938 (2012).

16. Zheng, X., Cahill, D. G., Krasnochtchekov, P., Averback, R. S. & Zhao, J. C. High-throughput thermal conductivity measurements of nickel solid solutions and the applicability of the Wiedemann-Franz law. *Acta Mater.* **55**, 5177–5185 (2007).

17. López-Honorato, E. *et al.* Thermal conductivity mapping of pyrolytic carbon and silicon carbide coatings on simulated fuel particles by time-domain thermoreflectance. *J. Nucl. Mater.* **378**, 35–39 (2008).

18. Cheaito, R. *et al.* Thermal conductivity measurements on suspended diamond membranes using picosecond and femtosecond time-domain thermoreflectance. in *Proceedings of the 16th InterSociety Conference on Thermal and Thermomechanical Phenomena in Electronic Systems, ITherm 2017* 706–710 (Institute of Electrical and Electronics

Engineers Inc., 2017). doi:10.1109/ITHERM.2017.7992555.

19. Grimm, D. *et al.* Thermal conductivity of mechanically joined semiconducting/metal nanomembrane superlattices. *Nano Lett.* **14**, 2387–2393 (2014).

20. Sood, A. *et al.* An electrochemical thermal transistor. *Nat. Commun.* **9**, 1–9 (2018).

21. Brown, D. B. *et al.* Spatial Mapping of Thermal Boundary Conductance at Metal-Molybdenum Diselenide Interfaces. *ACS Appl. Mater. Interfaces* **11**, 14418–14426 (2019).

22. Pek, E. K., Brethauer, J. & Cahill, D. G. High spatial resolution thermal conductivity mapping of SiC/SiC composites. *J. Nucl. Mater.* **542**, 152519 (2020).

23. Brethauer, J. & Cahill, D. G. *MAPPING THE THERMAL CONDUCTIVITY OF SIC/SIC COMPOSITES*. (2017).

24. O'Hara, K. E., Hu, X. & Cahill, D. G. Characterization of nanostructured metal films by picosecond acoustics and interferometry. *J. Appl. Phys.* **90**, 4852–4858 (2001).

25. Cahill, D. G., Goodson, K. & Majumdar, A. Thermometry and thermal transport in micro/nanoscale solid-state devices and structures. *J. Heat Transfer* **124**, 223–241 (2002).

26. Kang, K., Koh, Y. K., Chiritescu, C., Zheng, X. & Cahill, D. G. Two-tint pump-probe measurements using a femtosecond laser oscillator and sharp-edged optical filters. *Rev. Sci. Instrum.* **79**, 114901 (2008).

27. Cahill, D. G. *et al.* Nanoscale thermal transport. *Journal of Applied Physics* vol. 93 793–818 (2003).

28. Wei, C., Zheng, X., Cahill, D. G. & Zhao, J. C. Invited Article: Micron resolution spatially resolved measurement of heat capacity using dual-frequency time-domain thermoreflectance. *Rev. Sci. Instrum.* **84**, (2013).

29. *Technology for Polarization Measurement*. www.hindsintruments.com (2005).

30. Cahill, D. G. Analysis of heat flow in layered structures for time-domain thermoreflectance. *Rev. Sci. Instrum.* **75**, 5119–5122 (2004).

31. Rich, T. C. & Pinnow, D. A. Total optical attenuation in bulk fused silica. *Appl. Phys. Lett.* **20**, 264–266 (1972).

32. Zheng, Q. *et al.* Thermal conductivity of GaN, GaN 71, and SiC from 150 K to 850 K. *Phys. Rev. Mater.* **3**, 014601 (2019).

33. Cahill, D., Bullen, A., Pressures, L. S.-M.-H. T. H. & 2000, undefined. Interface thermal conductance and the thermal conductivity of multilayer thin films. *Citeseer*.

34. Tomko, J. A., Boris, D. R., Rosenberg, S. G., Walton, S. G. & Hopkins, P. E. Thermal conductance of aluminum oxy-fluoride passivation layers. *Appl. Phys. Lett.* **115**, 191901 (2019).

35. Cahill, D. G. & Watanabe, F. Thermal conductivity of isotopically pure and Ge-doped Si epitaxial layers from 300 to 550 K. *Phys. Rev. B - Condens. Matter Mater. Phys.* **70**, 1–3 (2004).

36. Costescu, R. M., Wall, M. A. & Cahill, D. G. Thermal conductance of epitaxial interfaces. *Phys. Rev. B - Condens. Matter Mater. Phys.* **67**, 054302 (2003).

37. *User's Manual Model SR844 RF Lock-In Amplifier*. www.thinkSRS.com (1997).

38. Katoh, Y. *et al.* Continuous SiC fiber, CVI SiC matrix composites for nuclear applications: Properties and irradiation effects. *J. Nucl. Mater.* **448**, 448–476 (2014).